\documentstyle[12pt,epsfig]{article}
\def\be {\begin{equation}}
\def\ee {\end{equation}}

\def\bea {\begin{eqnarray}}
\def\eea {\end{eqnarray}}

\begin{document}

\thispagestyle{empty}
\vskip 15pt

\begin{center}
{\Large {\bf Rapidity distribution of photons from an 
anisotropic {\em Quark-Gluon-Plasma} }}
\renewcommand{\thefootnote}{\alph{footnote}}

\hspace*{\fill}

\hspace*{\fill}

{ \tt{
Lusaka Bhattacharya\footnote{E-mail address:
lusaka.bhattacharya@saha.ac.in}
and Pradip Roy\footnote{E-mail address:
pradipk.roy@saha.ac.in}
}}\\

\small {\em Saha Institute of Nuclear Physics \\
1/AF Bidhannagar, Kolkata - 700064, INDIA}
\\

\vskip 40pt

{\bf ABSTRACT}
\end{center}

\vskip 0.5cm

We calculate rapidity distribution of photons due to
Compton and annihilation processes from 
{\em Quark Gluon Plasma} (QGP) with pre-equilibrium momentum-space 
anisotropy. We also include contributions from hadronic matter 
with late stage transverse expansion. A phenomenological model 
has been used for the time evolution of hard momentum scale 
$p_{\rm hard}(\tau)$ and anisotropy parameter $\xi(\tau)$. As 
a result of pre-equilibrium momentum-space anisotropy, we find 
significant modification of photons rapidity distribution. 
For example, with {\em fixed initial condition} (FIC) 
{\em free-streaming} ($\delta=2$) interpolating model we observe 
significant enhancement of photon rapidity distribution at 
fixed $p_T$, where as for FIC {\em collisionally-broadened} 
($\delta=2/3$) interpolating model the yield increases till 
$y \sim 1$. Beyond that suppression is observed. 
With {\em fixed final multiplicity} (FFM) {\em free-streaming} 
interpolating model we predict enhancement of photon yield which 
is less than the case of FIC. Suppression is always observed for 
FFM {\em collisionally-broadened} interpolating model.

\vskip 30pt

\section{Introduction}

Relativistic heavy ion colliders (RHIC), at Brookhaven National
Laboratory and upcoming Large Hadron Collider (LHC) at CERN, are 
designed to produce and study strongly interacting matter at 
high temperature and/or density. Experiments at the RHIC have 
already demonstrated that high $p_T$ hadrons in central $A+A$ 
collisions are significantly suppressed in comparison with that in 
binary scaled $p+p$ collisions~\cite{mustafa123}. This observation 
has been referred to as {\em jet-quenching} which clearly 
indicates towards the formation of 
{\em Quark Gluon Plasma} (QGP)\footnote 
{Theoretically QGP is expected to be formed when the temperature 
of nuclear matter is raised above its critical value, 
$T_c \sim 170$ MeV, or equivalently the energy density of 
nuclear matter is raised above $1~GeV/fm^3$} at the 
RHIC experiments. Another most important task is to 
characterize different properties of this new state of 
matter, such as isotropization/thermalization. 

The most difficult problem lies in the determination of 
isotropization and thermalization time scales 
($\tau_{\rm iso}$ and $\tau_{\rm therm}$)
\footnote{From now on we will concentrate on the most simplest 
possibility that is both the time-scale are the same, 
$\tau_{\rm therm}=\tau_{\rm iso}$.}. Studies on elliptic flow 
(upto  about $p_T \sim 1.5 - 2 $ GeV) using ideal hydrodynamics  
indicate that the matter produced in such collisions becomes 
isotropic with $\tau_{\rm iso} \sim 0.6$ fm/c~\cite{Pasi,Hirano,Tannenbaum}. 
On the contrary, perturbative estimates yield  much slower 
thermalization of QGP~\cite{PRC75_ref2}. However, recent hydrodynamical 
studies~\cite{0805.4552_ref4} have shown that 
due to the poor knowledge of the initial conditions, there is a
sizable amount of uncertainty in the estimation of thermalization or 
isotropization time. The other uncertain parameters are the 
transition temperature $T_c$, the spatial profile, 
and the effects of flow. Thus it is necessary to find suitable 
probes which are sensitive to these parameters. Electromagnetic 
probes have long been considered to be one of the most promising 
tools to characterize the initial state of the 
collisions~\cite{ann,jpr}. Because of the very nature of their 
interactions with the constituents of the system they tend to 
leave the system without much change of their energy and 
momentum. Photons (dilepton as well) can be one of such observables.

But, photons can carry information about the 
plasma initial conditions~\cite{janejpg,dksepjc,pasi_2} only if the 
observed flow effects from the late stages of the collisions can be 
understood and modeled properly. The observation of pronounced 
transverse flow in the photon transverse momentum distribution has 
been taken into account in model calculations of photon $p_T$ 
distribution at various beam 
energies~\cite{janejpg,dksepjc,clem,janerap,trenk-hep-ph/0408218}. 
It is found that because of the transverse kick the low energy 
photons populate the intermediate regime and consequently, the 
contribution from hadronic matter becomes comparable with that 
from the hadronic matter destroying the window where the 
contribution from QGP was supposed to dominate~\cite{trenk-hep-ph/0408218}. 
Apart from transverse flow effects, the investigation of 
longitudinal evolution using HBT correlation measurements has 
been done in Ref.~\cite{prc71064905}. It is shown that the 
decrease of $R_{\rm side}$ with $p_T$ ($> 2.5$ GeV) provides a 
good indication whether transverse flow is significant or not. But, 
so far as the pre-equilibrium emission is concerned this effect 
is not important. 
However, in the late stage, transverse expansion becomes important 
and we include this while estimating the photon yield from the 
hadronic matter.

Photon (dilepton) production from relativistic heavy-ion 
collisions has been extensively studied in 
Ref.~\cite{npa98,npa99, kap, turbide_prc72}. All these works are 
based  on the assumption of rapid thermalization of plasma with
$\tau_{\rm therm}=\tau_{\rm i}$, where $\tau_{\rm i}$ is the time 
scale of plasma formation. However, due to rapid longitudinal 
expansion of the plasma at early time this assumption seems 
to be very drastic because it ignores the momentum space 
anisotropy developed along the beam axis.

The phenomenological consequences of early stage pre-equilibrium 
momentum space anisotropy of QGP have been studied in 
Ref.~\cite{mauricio_prl,mauricio_prc} in the context of dileptons and 
in Ref.~\cite{lusaka_prc, lus_PRC2} in the context of photons. 
In Ref.~\cite{lusaka_prc} the effects of time-dependent 
momentum-space anisotropy of QGP on the medium photon 
production are discussed. It is shown that the introduction of 
early time momentum-space anisotropy can enhance the photon 
production yield significantly. Also the present authors calculate 
transverse momentum distribution of direct photons from various 
sources by taking into account the initial state momentum 
anisotropy of QGP and the late stage tranverse flow 
effects~\cite{lus_PRC2}. The total photon yield is compared with 
the recent measurement of photon transverse momentum distribution 
by the PHENIX Collaboration to extract the isotropization 
time~\cite{lus_PRC2}. It is found that the data can be reproduced 
with $\tau_{\rm iso}$ in the range $0.5 - 1.5$ fm/c. 
All these works show that the introduction of pre-equilibrium 
momentum space anisotropy has significant effect on the medium 
dilepton as well as photon productions. In the present work, we 
will be investigating the rapidity dependence of thermal photon 
in the presence of pre-equilibrium momentum space anisotropy. The 
importance of rapidity distributions of photons and dileptons 
produced in relativistic heavy ion collisions has been previously 
realized in Ref.~\cite{PRC51,PRD49,mauricio_epj}. As for example, 
fluctuations in the rapidity distributions can signal a 
phase transition, a supercooling of the fluid, or the presence 
of quark-matter bubbles, etc. Moreover, it was shown 
in Ref. \cite{prc71064905} that photon rapidity density can be a 
good probe to distinguish between Landau-like \cite{landau} and 
Bjorken-like \cite{jpg51} dynamics. Therefore, photon rapidity 
density should carry the signatures of pre-equilibrium momentum 
space anisotropy. The rapidity distribution of thermal photons 
produced in $Pb+Pb$ collisions at CERN SPS is also demonstrated 
in Ref.~\cite{dumitru} using a three-fluid hydrodynamical 
model. It is argued that rapidity dependence of photon spectra 
can provide cleaner insight about the rapidity dependence of 
the initial conditions, e. g. the temperature/time.

In absence of any precise knowledge about the dynamics at
early time of the collision, one can introduce phenomenological models
to describe the evolution of the pre-equilibrium phase. In this work,
we will use one such model, proposed in Ref.~\cite{mauricio_prl}, for
the time dependence of the anisotropy parameter, $\xi(\tau)$, and hard
momentum scale, $p_{\rm hard}(\tau)$. This model introduces four parameter
to parameterize the ignorance of pre-equilibrium dynamics: the parton
formation time ($\tau_i$), the isotropization time ($\tau_{\rm iso}$), which
is the time when the system starts to undergo ideal hydrodynamical
expansion and $\gamma$ sets the sharpness of the transition to
hydrodynamical behavior. The fourth parameter $\delta$ is introduced
to characterize the nature of pre-equilibrium anisotropy i.e whether
the pre-equilibrium phase is non-interacting or collisionally 
broadened. The phenomenological model in Ref.~\cite{mauricio_prl} 
assumes Bjorken's boost invariant expansion of the plasma. 
Therefore, this leads to a rapidity independent production 
of thermal photons. One can easily see that this can not be 
true for collisions involving nuclei having a finite energy. 
The experimental data are expected to be better described by 
rapidity dependent parton distribution functions. Therefore, 
in this work, we have used rapidity dependent quark 
and anti-quark distribution functions. The rapidity dependence 
of the distribution functions arises from the rapidity 
dependence of the initial temperature, $T_{\rm i}(\eta)$. 

The organization of the paper is as follows. In the next section
(section 2) we shall discuss the mechanisms of photon
production rate using an anisotropic phase space distribution along
with the space-time evolution of the matter. 
Results are presented in section 3 and finally we conclude in section 4.

\section{Formalism}

\subsection{Photon rate : Anisotropic QGP}

The lowest order mechanisms for photon emission from QGP are the
Compton ($q ({\bar q})\,g\,\rightarrow\,q ({\bar q})\,
\gamma$) and the annihilation ($q\,{\bar q}\,\rightarrow\,g\,\gamma$)
processes. 
The rate of photon production from 
anisotropic plasma due to Compton and annihilation processes has been 
calculated in Ref.~\cite{prd762007}. The soft contribution is 
calculated by evaluating the photon polarization tensor for an 
oblate momentum-space anisotropy of the system where the 
cut-off scale is fixed at $k_c \sim \sqrt g p_{\rm hard}$. 
Here $p_{\rm hard}$ is a hard-momentum scale that appears in the 
distribution functions.

The differential photon production rate for 
$1+2\to3+\gamma$ in an anisotropic medium is 
given by~\cite{prd762007}:
\begin{eqnarray} 
E\frac{dR}{d^3p}&=& \frac{{\mathcal{N}}}{2(2\pi)^3} 
\int \frac{d^3p_1}{2E_1(2\pi)^3}\frac{d^3p_2}{2E_2(2\pi)^3}
\frac{d^3p_3}{2E_3(2\pi)^3}
f_1({\bf{p_1}},p_{\rm hard}(\tau, \eta),\xi)f_2({\bf{p_2}},
p_{\rm hard}(\tau,\eta),\xi) \nonumber\\
&\times&(2\pi)^4\delta(p_1+p_2-p_3-p)|{\mathcal{M}}|^2 
[1\pm f_3({\bf{p_3}},p_{\rm hard}(\tau, \eta),\xi)]
\label{photonrate}
\end{eqnarray}
where, $|{\mathcal{M}}|^2$ represents the spin averaged matrix element
squared for one of those processes which contributes in the photon rate
and ${{\mathcal N}}$ is the degeneracy factor of the corresponding
process. $\xi$ is a parameter controlling the strength of the anisotropy 
with $\xi > -1$. $f_1$, $f_2$ and $f_3$ are the anisotropic 
distribution functions of the medium partons and will be 
discussed in the following. Here it is assumed that the infrared 
singularities can be shielded by the thermal masses for the 
participating partons. This is a good approximation at times short 
compared to the time scale when plasma instabilities start to play 
an important role. 

The anisotropic distribution function can be obtained~\cite{stricland} 
by squeezing or stretching an arbitrary isotropic distribution function 
along the preferred direction in the momentum space,
\begin{eqnarray}
f_{i}({\bf p},\xi, p_{\rm hard})=f_{i}^{\rm iso}(\sqrt{{\bf p}^{2}+\xi 
({\bf p.n})^{2}},p_{\rm hard}(\tau, \eta))
\label{dist_an}
\end{eqnarray}
where ${\bf n}$ is the direction of anisotropy. It is important to
notice that $\xi > 0$ corresponds to a contraction of the 
distribution function in the direction of 
anisotropy and $-1 < \xi < 0 $ corresponds to a stretching in the
direction of anisotropy. In the context of relativistic
heavy ion collisions, one can identify the direction of anisotropy with 
the beam axis along which the system expands initially. The hard momentum
scale $p_{\rm hard}$ is directly related to the average momentum of the 
partons. In the case of an isotropic QGP, $p_{\rm hard}$ can be identified 
with the plasma temperature ($T$).

\subsection{Photon rate: Hadronic matter}

Photons are also produced from different hadronic reactions from hadronic 
matter either formed initially (no QGP scenario) or realized as a result 
of a phase transition (assumed to be first order in the present work) 
from QGP. Photons from hadronic reactions and decays cannot 
be calculated in a model-independent way. The hadronic matter produced 
in heavy ion collisions is usually considered to be a gas of the low 
lying mesons $\pi$, $\rho$, $\omega$, $\eta$ and nucleons. Reactions 
between these as well as the decays of the $\rho$ and $\omega$ were 
considered to be the sources of thermal photons from hadronic 
matter~\cite{ann,kap,song}. 

We follow the calculations done in Ref.~\cite{turbide} where 
convenient parameterizations have been given for the reactions 
considered. These parameterizations will be used while doing the 
space-time evolution to calculate the photon yield from meson-meson 
reactions.

\subsection{Space time evolution}

The rate given in Eq.~(\ref{photonrate}) is the static rate which has to
be convoluted with the space-time history of the plasma to obtain
phenomenologically predictable quantities, for example, 
$dN/d^2 p_T dy$ for a given $p_T$ or $dN/d^2 p_T dy$ for a given $y$.
In our calculation, we assume an isotropic plasma is formed
at initial temperature $T_i$ and initial time (proper) $\tau_i$.
Subsequent rapid expansion of the matter along the longitudinal
direction causes faster cooling in the beam direction than in the
transverse direction. As a result the system becomes anisotropic
and remains so till $\tau = \tau_{\rm iso}$. 

The exact dynamics at the early-stage of the heavy ion collision is
almost unknown. Thus, a precise theoretical picture of the 
evolutions of $p_{\rm hard}$ and $\xi$ is not possible. 
However, we can always introduce phenomenological models 
to parameterize the ignorance. 
In this work, we shall closely follow the work of 
Ref.~\cite{mauricio_prc} to evaluate the rapidity 
distribution of photons from the first few Fermi of the 
plasma evolution. Three scenarios of the space-time evolution 
(as described in Ref.~\cite{mauricio_prc}) are the following: 
(i) $\tau_{\rm iso} = \tau_i$, the system evolves 
hydrodynamically so that $\xi =0$ and $p_{\rm hard}$ can be 
identified with the temperature ($T$) of the system 
(till date all the calculations have been performed in 
this scenario), (ii) $\tau_{\rm iso}\rightarrow \infty$, 
the system never comes to equilibrium, 
(iii) $\tau_{\rm iso} > \tau_i$ and $\tau_{\rm iso}$ is finite, 
one should devise a time evolution model for $\xi $ and 
$p_{\rm hard}$ which smoothly interpolates between 
pre-equilibrium anisotropy and hydrodynamics. We shall 
follow scenario (iii) (see Ref.~\cite{mauricio_prc} for 
details) in which case the time dependence of the 
anisotropy parameter $\xi$ is given by
\begin{eqnarray}
\xi(\tau,\delta) &=& (\frac{\tau}{\tau_i})^\delta-1 
\label{eq_xi}
\end{eqnarray}
where the exponent $\delta = 2~(2/3)$ corresponds to {\em free-streaming 
(collisionally-broadened)} pre-equilibrium momentum space anisotropy and
$\delta=0$ corresponds to complete isotropization. 
As in Ref.~\cite{mauricio_prc}, a transition width 
$\gamma^{-1}$ is introduced to take into account the smooth 
transition from non-zero value of $\delta$ 
to $\delta = 0$ at $\tau = \tau_{\rm iso}$. The time dependence of 
various quantities are, therefore, obtained in terms of a smeared step 
function \cite{mauricio_prl}:
\begin{equation} 
\lambda(\tau)=\frac{1}{2}(\tanh[\gamma(\tau-\tau_{\rm iso})/\tau_i]+1). 
\label{eq_lamda}
\end{equation}
For $\tau << \tau_{\rm iso} ( >> \tau_{\rm iso})$ we have 
$\lambda = 0 (1)$ which corresponds to {\em free-streaming} 
(hydrodynamics). With this, the time dependence of relevant 
quantities are as follows~\cite{mauricio_prc, mauricio_epj}:
\begin{eqnarray}
\xi(\tau,\delta) &=& \left(\frac{\tau}{\tau_i}\right)^
{\delta(1-\lambda(\tau))}-1,\nonumber\\
p_{\rm hard}(\tau, \eta) &=& T_i (\eta)~{\bar {\cal U}}^{1/3}(\tau),
\label{xirho}
\end{eqnarray}
where, 
\begin{eqnarray}
{\mathcal U}(\tau)&\equiv& \left[{\mathcal
    R}\left((\frac{\tau_{\rm iso}}{\tau})^\delta-1\right)
\right]^{3\lambda(\tau)/4}
\left(\frac{\tau_{\rm iso}}{\tau}\right)^{1-\delta(1-\lambda(\tau))/2},
\nonumber\\
{\bar {\cal U}}&\equiv& \frac{{\cal U}(\tau)}{{\bar {\cal U}}(\tau_i)},
\nonumber\\
{\mathcal R}(x)&=&\frac{1}{2}[1/(x+1)+\tan^{-1}{\sqrt
    {x}}/\sqrt{x}]
\label{utau}
\end{eqnarray}
and $T_i$ is the initial temperature of the plasma. 

To estimate the initial conditions we assume that i.e the 
longitudinal expansion approximately follows the
scaling law, $v_z= z/t $, we can relate the initial density to the
final multiplicity distribution by 
\begin{eqnarray}
n_i \tau_i= \frac{1}{\pi {R_{\perp}}^2}\frac{dN}{d\eta}
\end{eqnarray} 
and the initial temperature ($T_i$) is related with the initial
density by the following relation.
\begin{eqnarray}
n_i= \frac{\zeta (3)g_Q} {\pi^2}\ {T_i}^3
\end{eqnarray} 
where $g_Q$ is degeneracy factor. The multiplicity distribution 
can be parameterized as 
\begin{eqnarray}
\frac{dN}{d\eta}=\left(\frac{dN}{d\eta}\right)_0 \ 
\exp\left(-\frac{{\eta}^2}{2{\sigma}^2}\right)
\end{eqnarray}
where $(dN/d\eta)_0$ is the total multiplicity at $\eta=0$ and 
$\sigma=3(5)$ for RHIC (LHC) energies. 
The initial temperature is therefore a function of $\eta$, 
i. e $T_i=T_i(\eta)$.

%
%
\begin{figure}
\begin{center}
\epsfig{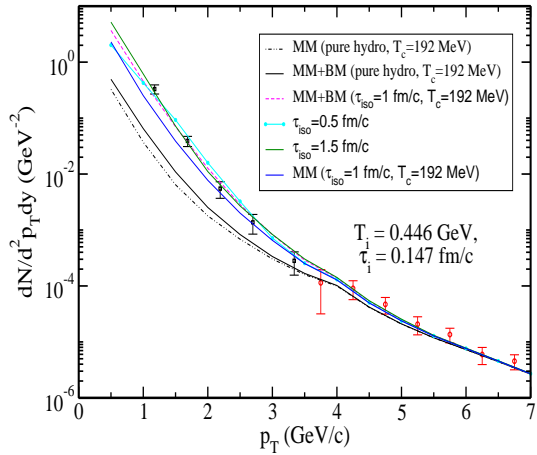}
\end{center}
\caption{(Color online)
Photon transverse momentum ($p_T$) distributions at RHIC energy with
initial condition $T_i=0.446$ GeV and $\tau_i=0.147$ fm/c. MM (BM) represents
the contribution from meson-meson (baryon-meson) interactions (see
\cite{lus_PRC2} for details).}
\label{fig3}
\end{figure}
%
%

So far, we have discussed the evolution during early 
stage only. For ($\tau > \tau_{\rm iso}$) 
we propose that $\tau_{\rm iso}$ onward the system is 
described by $(1+2)d$ ideal hydrodynamics. 
As the system becomes isotropic at 
$\tau = \tau_{\rm iso}$, $p_{\rm hard} (\tau_{\rm iso})$ 
and $\tau_{\rm iso}$ can be identified as the initial conditions, 
i. e., initial temperature and initial time for the hydrodynamic evolution. 
The initial conditions for ideal hydrodynamics is given by,
\bea
T_i^{\rm hydro}&=&p_{\rm hard} (\tau_{\rm iso})
\nonumber\\
 \tau_i^{\rm hydro}&=& \tau_{\rm iso}
\label{Tth}
\eea
Eventually the system undergoes a phase 
transition (assumed to be first order in the present work) at 
$\tau = \tau_f$, where $\tau_f$ is determined by the 
condition $p_{\rm hard}(\tau=\tau_f)=T_c $, where $T_c \sim 170$ MeV. 
The phase transition ends at $\tau_H = r_d \tau_f$,
where $r_d=g_Q/g_H$ is the ratio of the degrees of freedom in the two 
(QGP phase and hadronic phase) phases. 
Therefore, the total photon yield, arising from present scenario 
is given by the following equation, 
\begin{eqnarray}
\frac{dN}{d^2p_Tdy}=
\left[\int\,d^4x\, E\frac{dR}{d^3p}\right]_{\rm aniso} + 
\left[\int\,d^4x\, E\frac{dR}{d^3p}\right]_{\rm hydro}, 
\label{yield_total1} 
\end{eqnarray}
where the first term denotes the contribution from the anisotropic QGP 
phase and the second term represents the contributions evaluated in ideal 
hydrodynamics scenario. The rapidity density is defined as,
\begin{eqnarray}
\frac{dN}{dy} =  \int d^2p_T \frac{dN}{dyd^2p_T}  
\label{yield_total2} 
\end{eqnarray}
%

%
\begin{figure}
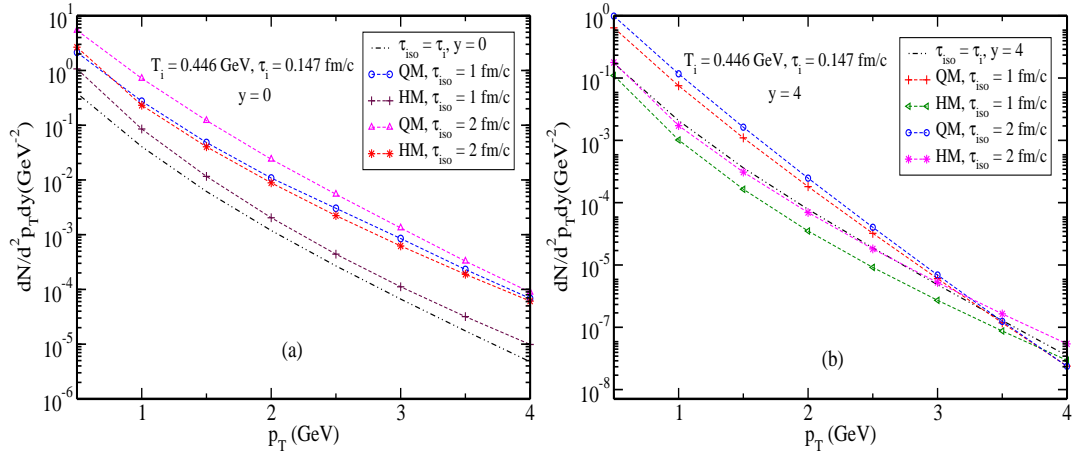

\begin{center}
\epsfig{file=T446_FIC_d2.eps,width=7cm,height=6cm,angle=0}
\epsfig{file=y4_T446_FIC_d2.eps,width=7cm,height=6cm,angle=0}
\end{center}
\caption{(Color online) The $p_T$ distribution of photons for FIC 
{\em free-streaming} ($\delta=2$) interpolating model at RHIC energy at 
(a) $y = 0$ and (b) $y = 4$.}
\label{ptdist_RH}
\end{figure}
 
In order to numerically compute the integrals in 
Eqs.~(\ref{yield_total1}) and~(\ref{yield_total2}), 
one needs to know the time dependence 
of $p_{\rm hard}$ and $\xi$ which has been discussed earlier.

After introducing the space-time evolution of relevant
quantities, we are now completely equipped for the
integration of Eqs. (\ref{yield_total1}) and~(\ref{yield_total2}). 
However, before going into the details of rapidity distributions 
of photons at the heavy ion collider experiments like RHIC and 
LHC, let us concentrate on few interesting features of the 
phenomenological model introduced for the evolution of 
$p_{\rm hard}(\tau,\eta)$ and $\xi(\tau)$. 

The simple model, introduced in section $2.3$, which smoothly 
interpolates between an initially non-equilibrium plasma to an 
isotropic plasma, is based on the assumption that the initial 
conditions are held fixed. The smooth interpolation 
(keeping the initial condition fixed) between anisotropic 
and isotropic phases (described in section $2$) results 
into a hard momentum scale (for $\tau>>\tau_{\rm iso}$) which is 
by a factor 
$[{\cal R}((\tau_{\rm iso}/\tau_i)^{\delta}-1)]^{0.25}$ 
larger compared to the momentum scale results from the 
hydrodynamic expansion of a system 
(with the same initial condition) from the beginning 
(see Eqs. (\ref{xirho}) and (\ref{utau})). As a consequence 
of this enhancement of $p_{\rm hard}(\tau,\eta)$, 
the {\em fixed initial condition} 
interpolating models (both {\em free-streaming} and 
{\em collisionally-broadened}) will result in generation of 
particle number during the transition from non-equilibrium to 
equilibrium phase. Moreover, the entropy generation increases 
with the increasing value of $\tau_{\rm iso}$. Thus, the requirement of 
bounded entropy generation can be used to put 
some upper bound on the value of $\tau_{\rm iso}$ for {\em fixed initial 
condition} interpolating models. As for example, if we allow maximum 
20\% entropy generation at RHIC, the maximum possible value 
of $\tau_{\rm iso}$ will be $1.2$ ($18$) fm/c for {\em fixed initial 
condition collisionally-broadened (free-streaming)} interpolating model. 

\begin{figure}
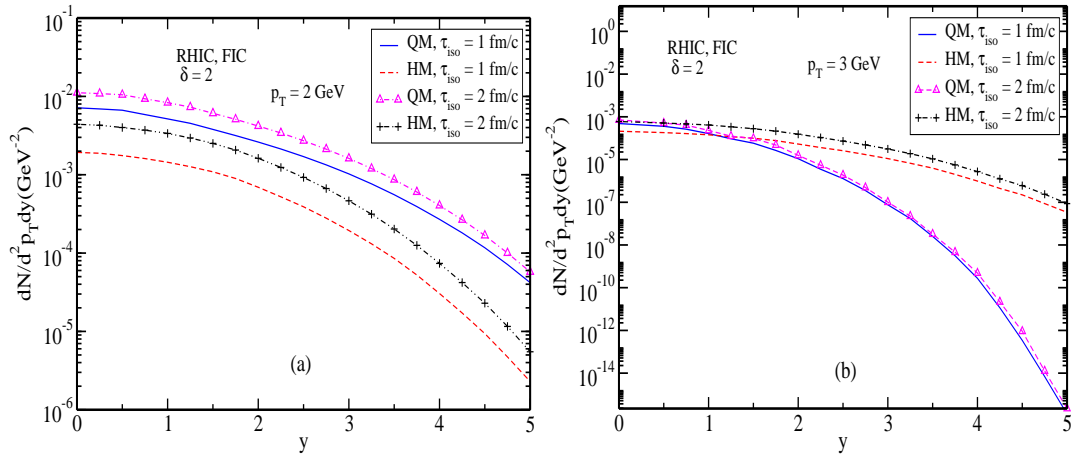

\begin{center}
\epsfig{file=QH_FICd2p2.eps,width=7cm,height=6cm,angle=0}
\epsfig{file=QH_FICd2p3.eps,width=7cm,height=6cm,angle=0}
\end{center}
\caption{(Color online) Rapidity distribution ($y$) of photon at 
RHIC energy from quark matter (QM) and hadronic matter (HM) 
for (a) $p_T = 2$ GeV and (b) $p_T = 3$ GeV.}
\label{ydist_RH}
\end{figure}

Due to the phenomenological constraints on the entropy generation, one
might not allow any entropy generation at all. In that case, one can
redefine $\bar{\cal U}(\tau)$ in Eq. (\ref{xirho}) to 
ensure {\em fixed final multiplicity} in this model. Since we 
know the amount of enhancement (which is respondable for this 
entropy generation) of $p_{\rm hard}$, the redefinition of 
$\bar{\cal U}(\tau)$ will be straight forward~\cite{mauricio_prc}:
\begin{eqnarray}
\bar{\cal U}(\tau)={\cal U(\tau)}
\left[{\cal R}((\tau_{\rm iso}/ \tau_{i} )^{\delta} - 1) \right]
^{-3/4}(\tau_{i}/\tau_{\rm iso})
\end{eqnarray}           
It is important to mention that this redefinition corresponds to a
lower initial temperature  (${p_{\rm hard}(\tau_i, \eta)}< T_i$) for
$\tau_{\rm iso}>\tau_i$. Larger value of isotropization time corresponds
to lower initial temperature.

\begin{figure}
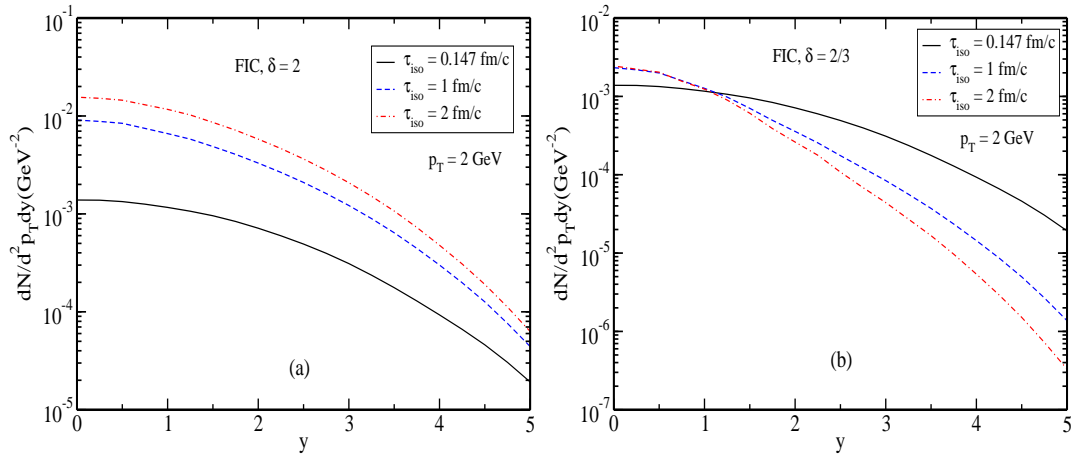

\begin{center}
\epsfig{file=FIC_d2.eps,width=7cm,height=6cm,angle=0}
\epsfig{file=FIC_d2_3.eps,width=7cm,height=6cm,angle=0}
\end{center}
\caption{(Color online) Photon rapidity distribution at RHIC energy 
for FIC interpolating model, at $p_T=2$ GeV for (a) $\delta=2$ and 
(b) $\delta=2/3$.}
\label{ficp2}
\end{figure}

\subsection{Photon rapidity distribution for fixed $p_T$}

The rapidity distribution of photons at fixed $p_T$ from a QGP 
or hadronic matter can be obtained by integrating 
Eq. (\ref{yield_total1}). We use a Monte-Carlo computer code to 
numerically evaluate Eq. (\ref{yield_total1}). 
For RHIC energies the formation time is taken as $\tau_i=0.147$ fm/c. 
This corresponds to an initial temperature of 
$T_i=446$ MeV at $\eta=0$ \cite{turbide_prc72}. For the 
LHC energies, we assume the formation time to be 
$\tau_i=0.073$ fm/c and the initial temperature to be 
$T_i=897$ MeV at $\eta=0$.

As a consequence of pre-equilibrium momentum-space anisotropy, 
significant modification of the thermal photon rapidity
distribution is expected. To quantify the effect of 
isotropization time on the rapidity distribution of the 
thermal photons, we define {\em photon modification} 
factor $\Phi(y,\tau_{\rm iso})|_{p_T}$ for fixed $p_T$ as 
the ratio of photon yields with and without 
pre-equilibrium momentum-space anisotropy, 
\begin{equation}
\Phi(y,\tau_{\rm iso})|_{p_T}=\left(\frac{dN(y,\tau_{\rm iso})}
{dyd^2p_T}\right)_{p_T}/\left(\frac{dN(y,\tau_{\rm iso}=\tau_i)}
{dyd^2p_T}\right)_{p_T}
\label{phi}
\end{equation}
The modification factor $\Phi$ is not measurable quantity. If 
there is an anisotropy in nature, the numerator can be 
measured (see Eq.~\ref{phi}). If there is no anisotropy in 
nature the denominator can be measured. But it is 
impossible to measure both simultaneously. However, $\Phi$ 
is a useful quantity only for demonstrative purpose.

\section{Results}

Before going to the description of photon rapidity distribution, 
we validate the present model in the context of PHENIX photon 
data~\cite{phenix}. It has already been demonstrated in 
Ref.~\cite{lus_PRC2}. We shall repeat here for completeness.  

To show the presence of initial state momentum anisotropy, 
we plot the total photon yield assuming hydrodynamic evolution 
from the very begining as well as with finite $\tau_{\rm iso}$ 
in Fig.~\ref{fig3}. 
It is clearly seen that some amount of anisotropy is needed to 
reproduce the data. We note that the value of $\tau_{\rm iso}$ needed 
to describe the data lies in the range 
$0.5$ fm/c$ \le \tau_{\rm iso} \le 1.5$ fm/c, independent of 
values of the transition temperatures~\cite{lus_PRC2}. 
It is to be noted that observables like photon $p_T$ distribution 
along with the ratio of photon to pion rapidity density will 
further strengthen the validity of the present model.

\subsection{Fixed initial condition (FIC) interpolating model}

%
\begin{figure}
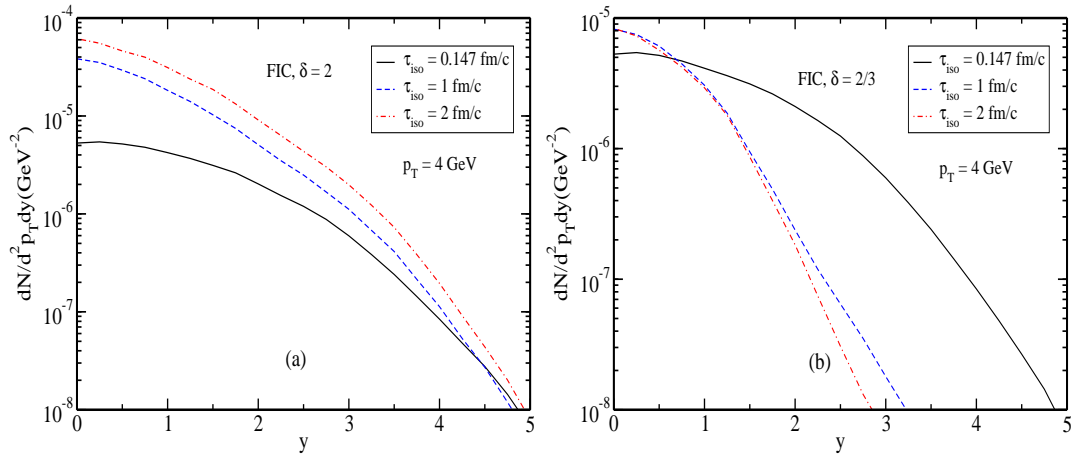

\begin{center}
\epsfig{file=pt4_FIC_d2.eps,width=7cm,height=6cm,angle=0}
\epsfig{file=pt4_FIC_d2_3.eps,width=7cm,height=6cm,angle=0}
\end{center}
\caption{(Color online) Photon rapidity distribution at RHIC energy 
for FIC interpolating model, at $p_T=4$ GeV for (a) $\delta=2$ 
and (b) $\delta=2/3$.}
\label{ficp4}
\end{figure}
%

Fixed initial condition (FIC) interpolating models always result into an
enhanced value of hard momentum scale as a consequence of pre-equilibrium
anisotropy. This feature of this model has already been discussed briefly 
before (see Ref.~\cite{mauricio_prc} 
for details). As a consequence of this enhancement 
of $p_{\rm hard}$, pre-equilibrium anisotropy increases the density 
of plasma partons with small rapidities. However, for 
higher rapidities, this enhancement is 
complemented by the suppression arising from the non-zero value 
of anisotropy parameter $\xi$. The anisotropic parton 
distribution function in Eq.~(\ref{dist_an}) clearly suggests 
that for the positive values of $\xi$, parton density 
decreases if we decrease the angle between the momentum of 
partons and the direction of anisotropy. From the very beginning, 
we have assumed that pre-equilibrium momentum-space anisotropy 
(in the heavy ion collisions) results from the rapid longitudinal 
expansion of the system immediately after the collision. 
Thus, in the context of relativistic heavy ion collision, 
one can identify the direction of anisotropy as the beam axis. 
Moreover, due to the rapid longitudinal cooling (as a consequence 
of rapid longitudinal expansion), one always finds oblate anisotropic 
distributions i.e positive value of $\xi$. Therefore, introduction of 
pre-equilibrium anisotropy with fixed initial condition increases the 
density of plasma partons moving in the transverse direction 
\cite{lusaka_prc} and at the same time decreases the density of 
plasma partons moving in the forward direction. This feature 
of fixed initial condition pre-equilibrium momentum-space 
anisotropy should be reflected in the photon 
rapidity distribution which we will discuss in the following.

%
%
\begin{figure}
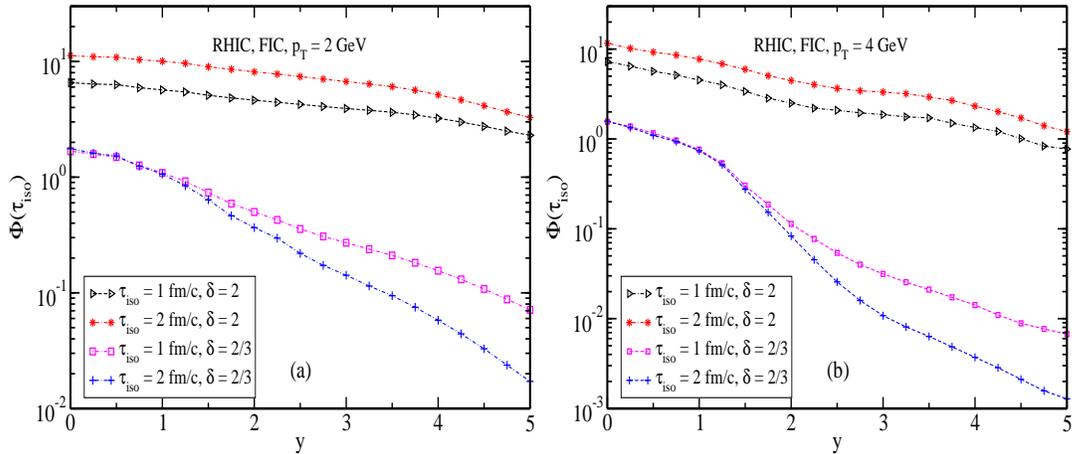

\begin{center}
\epsfig{file=pt2_FIC_phi.eps,width=7cm,height=6cm,angle=0}
\epsfig{file=pt4_FIC_phi.eps,width=7cm,height=6cm,angle=0}
\end{center}
\caption{(Color online) Modification factor for FIC interpolating model 
at RHIC energy, (a) $p_T=2$ GeV and (b) $p_T=4$ GeV.}
\label{ficphi}
\end{figure}
%
%

As a consequence of this enhancement of partons moving in the transverse 
direction and suppression of partons moving in forward direction, 
we expect an enhancement of the yield in the transverse direction 
($y=0$)\footnote{This feature was already established in 
our previous work \cite{lusaka_prc}.} and suppression in the 
longitudinal direction ($y \neq 0$).

To see the relative importance of the contributions from quark matter 
(QM) and hadronic matter (HM) we evaluate the $p_T$ distribution 
of photons for fixed rapidity from the evoluting fireball. 
Because the effect of tranverse flow is pronounced in 
the late stages of the collisions we neglect this effect in the early stage
and include it during the late stage. 
For $\tau \ge \tau_{\rm iso}$, the system is described by ideal 
relativistic hydrodynamics in $(1+2)d$~\cite{jpg50} with 
longitudinal boost invariance~\cite{jpg51} and cylindrical symmetry. 
Therefore, the total thermal photon yield for fixed $y$, 
arising from the present scenario (FIC interpolating model) 
is given by Eq.~(\ref{yield_total1}).

To elucidate the effect of transverse expansion, we first show results 
in the frame work of FIC {\em free-streaming} ($\delta=2$) 
interpolating model. In Fig.~\ref{ptdist_RH} we have 
plotted the $p_T$ distribution of photons for two 
different isotropization time 
($\tau_{\rm iso} = 1$ and $2$ fm/c) at (a) $y=0$ and (b) $y=4$. 
Both Figs.~\ref{ptdist_RH}a and~\ref{ptdist_RH}b clearly suggest 
that for $1\le p_T\le 4$ GeV, the 
photons from the quark matter (QM) dominates over the hadronic 
matter (HM). As we increase the isotropization time 
($\tau_{\rm iso}$) from $1$ to $2$ fm/c (see Fig. \ref{ptdist_RH}), 
both the contributions from quark matter and hadronic matter increases. 
However, quark matter contributions still dominate over hadronic 
matter contributions. 
Therefore, for the above mentioned $p_T$ range, 
the effects of pre-equilibrium momentum space anisotropy 
will be much more prominent irrespective of isotropization 
time ($\tau_{\rm iso}$). In support of this argument, we 
have plotted $dN/d^2{p_T}dy$ as a function of rapidity ($y$) 
for fixed $p_T$ ($p_T=2$ GeV) in Fig.~\ref{ydist_RH}a for 
two different values of isotropization time 
($\tau_{\rm iso}=1$ and $2$ fm/c respectively). 
Fig.~\ref{ydist_RH}a shows that the contributions from hadronic matter 
are always smaller than the contributions from the quark matter. 
However, for $p_T=3$ GeV, QM and HM contributions are of similar 
magnitude for $y \le 1 $ (see Fig.~\ref{ydist_RH}b). Beyond that the 
HM contributions dominates. 

The feature described above is better understood by 
looking at Figs. \ref{ficp2} and \ref{ficp4} where the total 
contributions have been plotted. In Fig. \ref{ficp2}, 
we have presented the rapidity distribution of photons ($dN/d^2p_Tdy$) 
as a function of rapidity ($y$) with fixed transverse momentum $p_T=2$ GeV 
(for three different values of isotropization time, 
$\tau_{\rm iso}=\tau_i,~1$ and $2$ fm/c) in the framework of 
{\em fixed initial condition} for (a) {\em free-streaming} 
($\delta=2$) and (b) {\em collisionally-broadened} 
($\delta=2/3$) interpolating models. The rapidity distribution 
of photons with $p_T=4$ GeV are
presented in Fig. \ref{ficp4} for (a) {\em free-streaming}
($\delta=2$) and (b) {\em collisionally-broadened} ($\delta=2/3$) 
interpolating model. 
Figs.~\ref{ficp2}a and \ref{ficp4}a show enhancements of 
photon yields (for the FIC {\em free-streaming} $(\delta=2)$) 
interpolating model) in the low rapidity region ($0 \le y \le 5$). 
Marginal suppressions in the higher rapidity 
region ($y \ge 5$) are observed. The situation is different for the 
FIC {\em collisionally broadened} $(\delta=2/3)$ interpolating model 
(see Figs.~\ref{ficp2}b and \ref{ficp4}b) which shows slight 
enhancements of photon yields in the low rapidity region 
($0 \le y \le 1.5$) and suppressions for the rest of the 
rapidity region ($y \ge 1.5$). 
This can be attributed to the fact that in the case of 
{\em collisionally-broadened interpolating} model we have included 
the possibility of momentum space broadening of the plasma partons 
due to interactions. As a consequence the hard momentum scale 
$p_{\rm hard}$ (which is related to the average momentum 
in the partonic distribution functions) decreases with time, whereas 
for {\em free-streaming} model, the hard momentum scale remains unchanged 
($p_{\rm hard}(\tau)=p_{\rm hard}(\tau_i)=T_i$, for $\tau<\tau_{\rm iso}$) 
upto $\tau=\tau_{\rm iso}$.

%
%
\begin{figure}
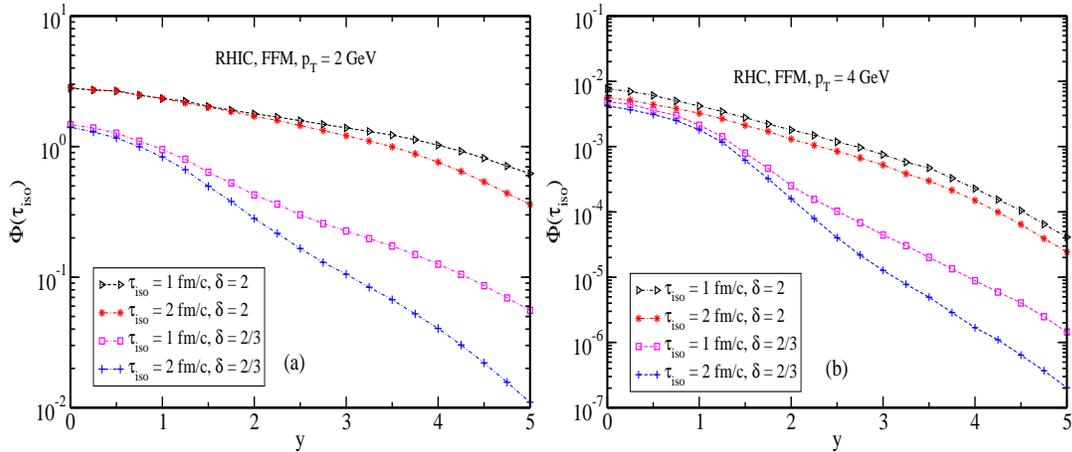

\begin{center}
\epsfig{file=pt2_FFM_phi.eps,width=7cm,height=6cm,angle=0}
\epsfig{file=pt4_FFM_phi.eps,width=7cm,height=6cm,angle=0}
\end{center}
\caption{(Color online) Modification factor for FFM interpolating model 
at RHIC energy, (a) $p_T=2$ GeV,(b) $p_T=4$ GeV.}
\label{ffmphi}
\end{figure}
%

These enhancement and suppression are more clearly revealed from 
Fig.~\ref{ficphi} which shows the {\em modification factors} 
for (a) $p_T=2$ GeV and (b) $p_T=4$ GeV. 
For a better comparison of {\em free-streaming} and 
{\em collisionally-broadened} pre-equilibrium phases, we 
have plotted the {\em modification factors} for $\delta=2$ and $2/3$ 
in the same graph.

We always find enhancement (one order of magnitude) for $\delta=2$. 
The trend of the graph shows that at larger rapidities 
$(y > 5 )$ the yield will be suppressed for the FIC 
{\em free-streaming} $(\delta=2)$ interpolating model. 
However, we observe suppression for FIC {\em collisionally-broadened}
interpolating model at rapidities $y \ge 1.5$.  
These findings are similar to that in Ref. \cite{lusaka_prc} where 
for $\delta=2$ with FIC considerable enhancement is observed at 
$y=0$. But for $\delta=2/3$, it is found that the enhancement 
is small.

\subsection{Fixed final multiplicity (FFM) interpolating model}

The problems, regarding the entropy generation (appears in {\em fixed
  initial condition} model), could be eliminated by enforcing {\em fixed
  final multiplicity}. However, enforcing {\em fixed final
  multiplicity} corresponds to a lower value of initial hard momentum
  scale ($p_{\rm hard}(\eta) < T_{i}(\eta)$). Moreover, in this case, the
  initial hard momentum scale will be isotropization time
  dependent i.e larger the value of $\tau_{\rm iso}$ lower will be the
  initial hard momentum scale. The suppression of hard momentum scale
  corresponds to a suppression in the plasma parton
  density compared to the {\em fixed initial condition} interpolating
model. Therefore, for {\em fixed final multiplicity} interpolating
  model, we predict more suppression in the photon yield compared to the FIC
  interpolating model.  
As a consequence of non-zero positive
  anisotropy parameter $\xi$, photons with higher rapidities will be
  more suppressed.

\begin{figure}
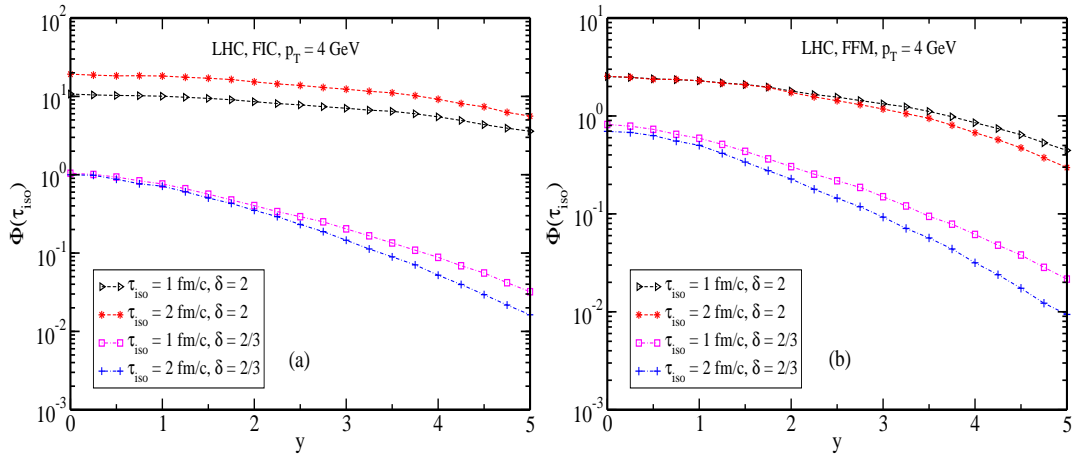

\begin{center}
\epsfig{file=LHC_pt4_FIC_phi.eps,width=7cm,height=6cm,angle=0}
\epsfig{file=LHC_pt4_FFM_phi.eps,width=7cm,height=6cm,angle=0}
\end{center}
\caption{(Color online) Modification factors, 
(a) FIC and (b) FFM interpolating models at LHC energy at $p_T=4$ GeV.}
\label{LHC_pt4}
\end{figure}

These features (compared to {\em fixed initial condition} 
interpolating model) can be better understood by looking at the 
{\em modification factors} for {\em fixed final multiplicity} 
interpolating model displayed in Fig.~\ref{ffmphi} for (a) $p_T=2$ GeV and (b)
$p_T=4$ GeV at RHIC energy. 
For $\delta =2$ (see Fig.~\ref{ffmphi}a) 
enhancement is observed for the rapidity range upto $y\le 4$ GeV. 
After that ($y\ge 4$), we observe the suppression of the photon yield. 
But for $\delta=2/3$ (see the same Fig), we observe slight enhancement of the
photon yield upto $y \le 1.5$ and suppression for the rest of the 
rapidity region. 
For $\delta=2/3$ at lower rapidities the rate is close to 
unity and as we go to higher rapidities the rate is reduced.   
It is to noted that although, in the case of FFM interpolating 
model, we have started with a lower value of initial hard momentum 
scale ($p_{\rm hard}(\tau_i) \le T_i (\tau_i)$), we obtain an 
enhancement in the low rapidity region (Fig.~\ref{ffmphi}a) 
at $p_T=2$ GeV  for both $\delta=2$ and $2/3$ respectively. For 
higher $p_T$ ($4$ GeV) the yield is suppressed at all rapidities 
irrespective of the values of $\delta$ (Fig.~\ref{ffmphi}b). 
However, the amount of suppression, for $\delta=2/3$ at rapidities
$y \ge 2$, is large compared to the {\em free-streaming} 
pre-equilibrium phase ($\delta=2$). 
For a more realistic case ($\delta=2/3$) the suppression is by a 
factor of $0.85$ which is in accordance with Ref. \cite{mauricio_epj} for 
the case of dilepton.
 
In Fig.~\ref{LHC_pt4}, we have presented the {\em photon modification} 
factor for $p_T=4$ GeV as a function of photon rapidity for 
$\sqrt{s_{NN}}=5.5$ TeV. 
Fig.~\ref{LHC_pt4}a shows enhancement (suppression) for {\em free-streaming} 
({\em collisionally-broadened}) pre-equilibrium phase for FIC. 
On the other hand, for FFM model (see Fig.~\ref{LHC_pt4}b), we obtain 
suppression for {\em collisionally-broadened} pre-equilibrium phase. 
For {\em free-streaming} pre-equilibrium phase, we find small 
enhancement in the lower rapidity region and suppression for 
higher rapidities.

\subsection{Photon rapidity density}

There is one more phenomenologically interesting observable namely the 
photon rapidity density, $dN /dy$, which is defined in
Eq.~(\ref{yield_total2}).  
Its shape at least is not expected to be affected by the transverse 
expansion of the system. Therefore, this quantity is 
phenomenologically very relevant. We have assumed two 
different values of isotropization time 
($\tau_{\rm iso}=1, 2$ fm/c) to compute the rapidity 
density of photons from an anisotropic QGP. Since the 
detection of very soft photons is experimentally challenging, 
photons having their transverse momentum greater than $1$ GeV is 
considered. 

The consequences of pre-equilibrium momentum-space anisotropy (with
{\em fixed initial condition} and {\em fixed final multiplicity}) 
on the photon production have already been discussed in 
section $3.1$ and $3.2$ in details.

%
\begin{figure}
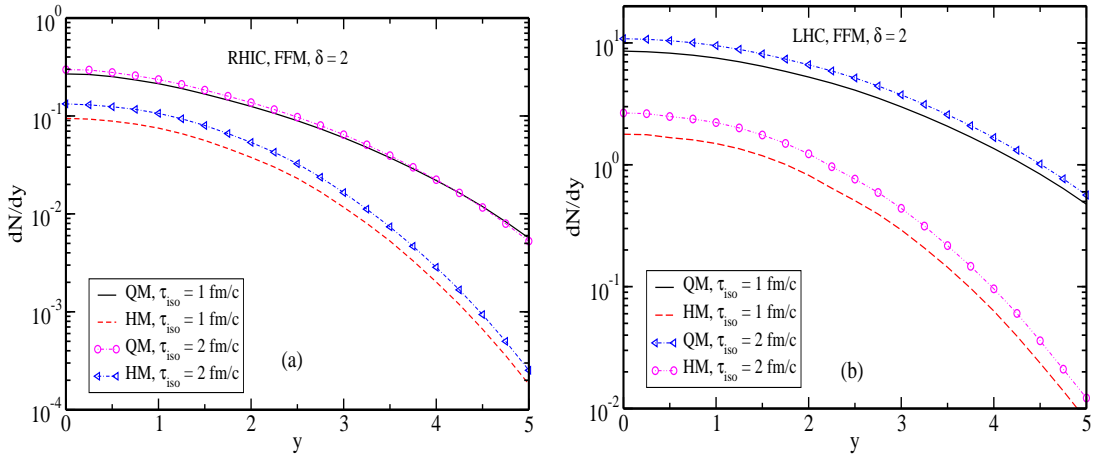

\begin{center}
\epsfig{file=DNDY_RH_FFM_d2.eps,width=7cm,height=6cm,angle=0}~~
\epsfig{file=LHC_DNDY_FFM_d2.eps,width=7cm,height=6cm,angle=0}
\end{center}
\caption{(Color online) Photon rapidity density 
at (a) RHIC and (b) LHC energies for FFM {\em free-streaming} 
($\delta=2$) interpolating models.}
\label{dndyffmd2}
\end{figure}


We show the effects of tranverse expansion to the rapidity density in 
Fig.~\ref{dndyffmd2}, where we have plotted the individual 
contributions of photon rapidity density from the quark matter (QM) 
and the hadronic matter (HM) in the frame work of 
{\em free-streaming} ($\delta=2$) FFM interpolating model at 
(a) RHIC and (b) LHC energies. As we vary $\tau_{\rm iso}$ from 
$1$ to $2$ fm/c (in both the Figs.~\ref{dndyffmd2}a 
and \ref{dndyffmd2}b), the individual contributions from the 
quark matter and the hadronic matter increase. At the same time 
rapidity density of photons from the quark matter is higher than 
the contributions from the hadronic matter. This is true for the 
RHIC as well as LHC energies (see Fig.~\ref{dndyffmd2}). This result 
is consistent with Fig.~\ref{ydist_RH}.

Let us now mention the main features of our observations. 
With FIC for both values of $\delta$ we see moderate 
enhancement at not too large rapidities. Similar feature has 
been noted earlier~\cite{lusaka_prc}. This can be attributed 
to the fact that momentum anisotropy enhances the density 
of plasma partons in the tranverse direction. However, it is 
also to be noted that for larger rapidities the rate is 
suppressed for both the values of $\delta$. This is due to 
the fact that the density of partons with higher longitudinal 
momentum (larger $y$) decreases. 
It is worthwhile to mention that the effects of pre-equilibrium 
momentum space anisotropy for $\delta=2$ is more compared 
to that for $\delta=2/3$. This is because $\delta=2/3$ 
corresponds to close to isotropization. With FFM we observe 
marginal enhancement at low rapidities but considerable 
suppression at higher rapidities for both {\em free-streaming} and 
{\em collisionally-broadened} interpolating models. 
This suppression can be described in 
two ways. Due to rapid longitudinal expansion the 
distribution function becomes anisotropic. Photons with the 
larger values of longitudinal momentum are reduced compared 
with the photons with isotropic distribution function. Maximum 
amount of momentum-space anisotropy achieved in the early 
times will be the important cause of the suppression. The 
suppression will also depend on the time dependence of the 
anisotropy parameter $\xi$. Another source of rapidity 
dependence is given by hard momentum scale ($p_{\rm hard}$) which 
depends on the initial temperature ($T_i$) and hence on 
rapidity. So we see that the hard momentum scale 
is directly related with the $\eta$ even in the case of 
instantaneous thermalization and satisfies the relation 
in Eq.~(\ref{xirho}).

It is important to mention that photon rapidity density 
significantly depends on the expansion dynamics of the system. In 
absence of any theoretical knowledge, one can introduce different 
models for the expansion dynamics of the system like, 
Bjorken and Landau dynamics etc. In Ref.~\cite{prc71064905}, it is 
shown that different expansion scenarios 
predicts different shape for photon rapidity density and thus, the 
shape of photon rapidity density can be used to distinguish 
different expansion scenarios. It is argued that the actual 
expansion scenario lies between Bjorken and Landau hydrodynamics 
by using various photonic observables~\cite{prc71064905}.

\begin{figure}
\begin{center}
\epsfig{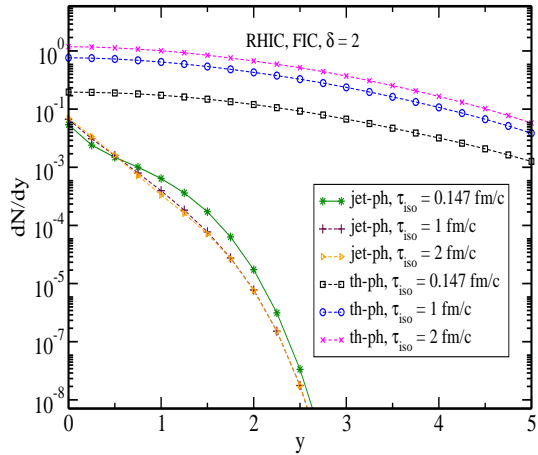}
\end{center}
\caption{(Color online) thermal photon and jet-photon rapidity density 
at RHIC energy for FIC {\em free-streaming} ($\delta=2$) 
interpolating model.}
\label{jet_rap}
\end{figure}

In this article, we have only concentrated on the thermal photon 
rapidity (from both the quark matter and the hadronic matter) 
density as a probe of pre-equilibrium anisotropy. However, 
there are other non-thermal sources of photon like, hard photons, 
fragmentation photons, decay photons etc. Hard photons are 
produced in the initial hard scattering of colliding nuclei and are 
insensitive to the later stage evolution of the QGP. Therefore, they 
do not carry any information about the pre-equilibrium anisotropy. 
Similarly, fragmentation and decay photons are also not sensitive to 
the pre-equilibrium phase. 
It is to be mentioned that apart from thermal photons from quark 
matter, thermal photons are also produced from hot hadronic matter 
where the late stage transverse expansion plays important role. 
In this work we include this effects of late stage transverse expansion. 
There is another important source of photons namely, the photons 
from jet-plasma interaction. The importance of jet-plasma interaction 
in the context of PHENIX \cite{phenix} photon data was shown in 
Ref.~\cite{turbide_prc72}. For jet-plasma interaction, the 
jet-parton distribution functions do not depend on the 
pre-equilibrium phase. However, the plasma parton densities 
depend on the pre-equilibrium anisotropy. Therefore, the jet-plasma 
interaction is sensitive to the pre-equilibrium anisotropy. However, 
the sensitivity is smaller compared to the thermal photons. 
Because, for thermal photons, both the initial state 
partons are sensitive to the anisotropy. Moreover, the 
jet-photon contribution dominates over the thermal photon 
contribution only in the high $p_T$ region. Therefore, photons from 
jet-plasma interaction insignificantly contributes to the photon 
rapidity density in the $p_T$ range considered here.
 
In Fig.~\ref{jet_rap}, we have plotted the contribution from jet plasma
interaction along with the anisotropic contributions for two values of
isotropization times in the frame work of FIC {\em free-streaming}
interpolating model. We have also presented the contribution from isotropic
QGP (i.e. corresponding thermal photon rapidity density) in the same 
figure for comparison. It is seen that the contribution from 
jet-plasma interaction is well below the anisotropic contribution, 
leaving behind a window - where the effects of anisotropy could be seen. 
Fig. \ref{jet_rap} clearly shows that jet-photon contribution is not 
very significant in the context of thermal photon rapidity density.


\section{Conclusion}

To summarize, we have investigated the effects of the pre-equilibrium
momentum space anisotropy of the QGP along with the effects of 
late stage tranverse expansion on the $p_T(y)$ distribution at 
fixed $y(p_T)$ and rapidity density ($dN/dy$) of photons. To
describe space-time evolution of hard momentum scale, $p_{\rm hard}(\tau)$
and anisotropy parameter, $\xi(\tau)$, phenomenological 
models have been used~\cite{mauricio_prc}. These
phenomenological models assume the existence of an intermediate time
scale called the isotropization time ($\tau_{\rm iso}$). The first model is
based on the assumption of fixed initial condition. However, enforcing fixed
initial condition causes entropy generation. Therefore, we have 
also considered another model, which assumes the fixed final 
multiplicity. Both the possibilities of {\em free-streaming} 
and {\em collisionally-broadened} pre-equilibrium phase of the 
QGP are considered. The rapidity distribution of 
photons for different isotropization times in the frame work of 
these phenomenological models have been estimated.
We observed that, for fixed initial condition, a {\em free streaming} 
interpolating model can enhance the photon yield 
significantly for rapidities upto $y \sim 4.5$. 
However, for {\em collisionally-broadened} pre-equilibrium 
phase with fixed initial condition, the enhancement of photon 
yield is upto $y \sim 1.5$. After that we observe the suppression of the 
photon yield for the entire rapidity region ($y \ge 1.5$). 
Since fixing the final multiplicity 
reduces the initial hard momentum scale or equivalently the 
initial energy density, we observe slight enhancement in the low rapidity
region and significant suppression (both for the {\em free-streaming} 
and {\em collisionally-broadened} interpolating models) for the rest of the
rapidity region. This suppression can be explained as a 
consequence of the combined effect of the anisotropy in 
momentum-space achieved at early times due to
expansion and the rapidity dependence of the hard momentum scale. 
For RHIC energies at $p_T=2$ GeV, QM contribution dominates over the HM 
contribution. However, for $p_T=3$ Gev, the later dominates over QM for $y \le
1$. But as far as the total contribution is concerned, we always find 
significant modification (enhancement or suppression depending upon the 
initial conditions used) of the yield in presence of pre-equilibrium 
momentum space anisotropy.
 
The other observables like heavy-quark transport~\cite{subhra}, 
jet-medium-induced electromagnetic and gluonic radiation could be 
phenomenologically very useful in order to detect the consequences of
pre-equilibrium momentum-space anisotropy. \\   

{\bf Acknowledgments:} We would like to thank D. K. Srivastava and 
S. Sarkar for their useful suggestions.

\end{document}